\PassOptionsToPackage{table}{xcolor}
\documentclass[sigconf, nonacm]{acmart}
\usepackage{amsmath, amsfonts}
\usepackage{algorithmic}
\usepackage{graphicx}
\usepackage{textcomp}
\usepackage{enumitem}
\usepackage{multirow}
\usepackage{listings}
\usepackage{subcaption}
\usepackage{caption}
\usepackage{cleveref}
\usepackage[normalem]{ulem}

\fancypagestyle{plain}{%
  \fancyhf{} 

}

\acmConference{}{}{}
\acmYear{}
\copyrightyear{}

\setcopyright{none}
\renewcommand\footnotetextcopyrightpermission[1]{}
\definecolor{diffgreen}{RGB}{230, 255, 230}
\definecolor{diffred}{RGB}{255, 230, 230}
\definecolor{diffgreentxt}{RGB}{34, 100, 34}
\definecolor{diffredtxt}{RGB}{160, 30, 30}
\definecolor{codebg}{RGB}{250, 250, 250}
\definecolor{keywordblue}{RGB}{0, 0, 170}
\definecolor{commentgray}{RGB}{110, 110, 110}
\definecolor{linenumgray}{RGB}{160, 160, 160}

\lstdefinestyle{diffleft}{
    language=C++,
    basicstyle=\ttfamily\tiny,
    backgroundcolor=\color{codebg},
    frame=single,
    framerule=0.3pt,
    rulecolor=\color{black!15},
    xleftmargin=14pt,
    xrightmargin=1pt,
    framexleftmargin=14pt,
    breakatwhitespace=false,
    columns=flexible,
    keepspaces=true,
    showstringspaces=false,
    tabsize=2,
    keywordstyle=\color{keywordblue}\bfseries,
    commentstyle=\color{commentgray}\itshape,
    stringstyle=\color{purple},
    escapechar=|,
    aboveskip=0pt,
    belowskip=0pt,
    numbers=left,
    numberstyle=\tiny\color{linenumgray},
    numbersep=4pt,
    stepnumber=1,
}

\lstdefinestyle{diffright}{
    language=C++,
    basicstyle=\ttfamily\tiny,
    backgroundcolor=\color{codebg},
    frame=single,
    framerule=0.3pt,
    rulecolor=\color{black!15},
    xleftmargin=1pt,
    xrightmargin=1pt,
    framexleftmargin=1pt,
    breakatwhitespace=false,
    columns=flexible,
    keepspaces=true,
    showstringspaces=false,
    tabsize=2,
    keywordstyle=\color{keywordblue}\bfseries,
    commentstyle=\color{commentgray}\itshape,
    stringstyle=\color{purple},
    escapechar=|,
    aboveskip=0pt,
    belowskip=0pt,
    numbers=none,
}

\title{GR-Evolve: Design-Adaptive Global Routing via LLM-Driven Algorithm Evolution}

\author{Taizun Jafri}
\affiliation{%
  \institution{Arizona State University}
  \city{Tempe}
  \state{Arizona}
  \country{US}
}
\email{tsjafri@asu.edu}

\author{Vidya A. Chhabria}
\affiliation{%
  \institution{Arizona State University}
  \city{Tempe}
  \state{Arizona}
  \country{US}
}
\email{vachhabr@asu.edu}

\begin{document}

\begin{abstract}
Modern ASIC design is becoming increasingly complex, driving up design costs while limiting productivity gains from existing EDA tools. Despite decades of progress, current tools rely on fixed heuristics and expose limited control through tool hyperparameters that require extensive manual tuning to achieve acceptable quality of results (QoR). While prior work has explored learning-based optimization and design-specific hyperparameter tuning, these approaches operate within the constraints of static tool algorithm implementations and do not adapt the underlying algorithms to individual designs. To address this limitation, we introduce the concept of design-adaptive EDA tooling, in which the internal algorithms of EDA tools are automatically specialized to the characteristics of a given design. We instantiate this paradigm through GR-Evolve, a code evolution framework that leverages an agentic large language model (LLM) to iteratively modify global routing source code using QoR-driven feedback. The framework equips the LLM with persistent contextual knowledge of open-source global routers along with an integrated toolchain for QoR evaluation within the OpenROAD infrastructure. We evaluate GR-Evolve across seven benchmark designs across three technology nodes and demonstrate up to 8.72\% reduction in post-detailed-routing wirelength over existing baseline routers, highlighting the potential of LLM-driven EDA code evolution for design-adaptive global routing.
\end{abstract}

\maketitle

\section{Introduction} \label{sec:introduction}

\noindent
{\bf Motivation.} Global routing is a critical stage in the physical design flow of integrated circuits, where approximate routing paths are determined for all nets across the chip. During this phase, the routing region is partitioned into a grid of global routing cells, called GCells, and each net is assigned a sequence of GCells through which it will pass, without specifying the exact locations of the wires. The primary objective is to efficiently allocate routing resources while minimizing congestion, wirelength, and delay. Global routing directly impacts the routability and performance of the final design. As modern chip designs continue to scale in complexity, effective global routing has become increasingly important to design closure. 

\begin{figure}[h]
    \centering
    \includegraphics[width=0.4\textwidth]{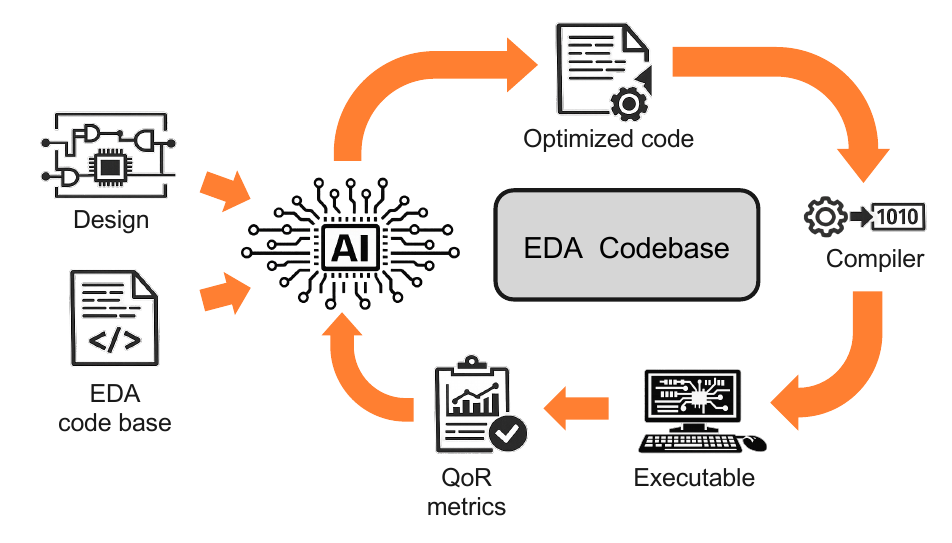}
    \caption{Design-adaptive EDA tools via LLM-driven EDA tool source code generation.}
    \label{fig:design-adaptive-eda}
    \vspace{-5mm}
\end{figure}

\noindent
{\bf Prior global routing work.} Modern academic global routers~\cite{fastroute, CUGR, SPRoute} employ diverse algorithmic frameworks to address large-scale routing challenges. FastRoute~\cite{fastroute} employs an efficient rip-up-and-reroute scheme that combines congestion-driven Steiner tree construction with 3-bend routing to achieve fast convergence. CUGR~\cite{CUGR} operates directly in 3D space, leveraging 3D pattern routing and multi-level maze routing guided by a probabilistic resource model to improve detailed routability. SPRoute~\cite{SPRoute} introduces a soft-capacity mechanism that reserves routing resources based on local pin density and employs parallelization for runtime efficiency.

Despite these advances, all existing global routers rely heavily on human-designed heuristics and fixed cost functions to explore the routing solution space. As a result, their performance is inherently design-dependent, and no single router consistently achieves the best quality of results (QoR) and runtime across all benchmarks. This highlights a fundamental limitation of current EDA practice: the use of static tools that are applied uniformly across diverse designs (one-size fits all). While one could attempt to tune hyperparameters within a given tool, these approaches remain limited. They do not enable fine-grained adaptation of the underlying algorithms, nor do they allow discovery of new heuristics tailored to specific design characteristics. Consequently, a significant portion of achievable QoR remains unexplored. To the best of our knowledge no prior global routing work has explored LLM-driven source code evolution to automatically generate design-adaptive routing algorithms, leaving this opportunity largely untapped.  In this work, we address this limitation by recasting global routing as a program optimization problem. Instead of optimizing routing solutions or tuning parameters of a fixed tool, we seek to optimize the routing algorithm itself. Specifically, given a baseline global router and a target design, the goal is to automatically evolve the router’s source code to produce a design-specialized algorithm that improves QoR under full routing flows. This formulation shifts the focus from design space exploration to \textit{design–tool co-exploration}, where both the design and the underlying EDA tool are jointly optimized.

\noindent
\textbf{Opportunity.} Large language models (LLMs) have undergone rapid evolution in the broader artificial intelligence (AI) community.  LLMs have been integrated into agentic systems that interact directly with external tools~\cite{codex, claudecode}, execute actions, observe outcomes, and iteratively refine behavior through feedback. This idea was advanced by AlphaEvolve~\cite{alphaevolve}, which integrates LLMs into large-scale evolutionary optimization pipelines that operate over entire codebases and algorithmic systems. By coupling LLM-driven proposal generation with automated evaluation, selection, and iteration, AlphaEvolve marks a transition from LLMs as interactive assistants to LLMs as autonomous components. These advances have been adopted in some early works for physical design and SAT. AuDoPEDA~\cite{AuDoPEDA} formalizes a closed-loop autonomous coding pipeline that systematically improves QoR for the open-source OpenROAD EDA stack by combining documentation, planning, and execution. However, this work focuses on placement and timing optimization and does not modify the global router. Another recent work~\cite{yu-sat}, inspired by AlphaEvolve, has also demonstrated the feasibility of extending LLM-driven code evolution to full repository scale. SATLUTION~\cite{yu-sat} is an example of such a framework, applying LLM-based evolutionary optimization to Boolean satisfiability.

\noindent
\textbf{Our work.} In this work, we recast global routing as a program optimization problem, where the objective is to automatically evolve the source code of a routing algorithm to specialize it for a target design. We introduce the concept of \emph{design-adaptive EDA tools}, enabled by LLMs that generate and evolve repository-scale global routing code to improve QoR through \emph{design--tool co-exploration}. Fig.~\ref{fig:design-adaptive-eda} illustrates this paradigm. We instantiate this concept in the context of global routing. We present GR-Evolve, an agentic framework that leverages an LLM-powered command-line interface~\cite{codex} to iteratively evolve global router source code, automatically specializing routing heuristics and cost functions within the router implementation to the characteristics of a given design.  The framework provides the agent with persistent contextual knowledge, including three open-source global router codebases, concise summaries of their routing strategies, and an integrated toolchain for building \cite{OpenROAD_firstLearnings} and executing the OpenROAD-flow-scripts (ORFS) on placed designs. In a closed-loop cycle, the agent reasons about routing heuristics, modifies source code, compiles the updated router within OpenROAD, and evaluates its performance on a target design.  Through successive iterations, the framework synthesizes specialized global routers tailored to patterns of individual designs. GR-Evolve performs a bounded exploration of the program space and selects the best-performing router observed during the search, rather than attempting to converge to a single optimal solution. We validate this approach on seven benchmark designs, ranging from 15k to 270k nets, across three technology nodes, generating variants from three baseline routers. Our contributions are:

\begin{itemize}[nosep, leftmargin=*]
\item We are the first to formulate global routing as a program-level optimization problem over source code.
\item We propose a design-adaptive agentic framework that automatically evolves global routing algorithms on a per-design (design-tool co-exploration) basis using LLM-driven code modification.
\item We develop a stateless, version-controlled evolution framework that externalizes all state into persistent artifacts, enabling scalable long-horizon optimization.
\item We formulate router evolution as a multi-objective optimization problem that reduces detailed routing (DR) wirelength (WL) and via count (VC), while maintaining awareness of the global routing (GR) runtime (RT).
\item We evaluate the impact of global routers using post-DR metrics.
\item We analyze diffs between the baseline router and the evolved router code to extract insights into design-specific characteristics captured by the learned modifications.
\end{itemize}
\noindent
Our GR-Evolve framework has been open-sourced on this GitHub repository~\cite{gr-evolve-gh}.

\section{Background and Prior Work}

\noindent
\textbf{Agentic Work for LLMs.}
The concept of LLMs as agents was formalized by ReAct\cite{react-test}, which proposed interleaving reasoning traces and external actions within a unified decision loop. By augmenting the action space of an LLM to include both executable actions and internal “thought” steps, ReAct enables planning and environment interaction through a closed feedback mechanism. This reasoning–action paradigm established the foundation for modern agentic LLM systems. Extending this paradigm to algorithm discovery, AlphaEvolve\cite{alphaevolve, Openevolve} introduces an evolutionary coding agent that iteratively improves executable programs through LLM-generated code mutations guided by automated evaluation. Rather than searching directly over candidate solutions, AlphaEvolve evolves the underlying algorithms themselves, demonstrating that LLM-guided evolutionary refinement can surpass human-designed methods in mathematical and computational domains.  

In parallel, tool-augmented agentic systems extend a single state-of-the-art LLM with execution capabilities, including code execution, file manipulation, and command-line interaction. Commercial LLM vendors, including OpenAI\cite{OpenAI_ChatGPT}, Anthropic\cite{Anthropic_Claude}, and Google\cite{Google_Gemini_CLI}, have developed agents that provide models with structured tool access to operate in real development environments, thereby enabling iterative debugging, testing, and refinement\cite{OpenAI_Codex, Anthropic_Claude, Google_Gemini_CLI}. Collectively, these systems demonstrate that coupling reasoning with execution feedback transforms LLMs from static text generators into autonomous, task-oriented agents.

\noindent
\textbf{Agentic AI for Physical Design.}
Recent work \cite{ORFS-Agent,openroad-agent-asu} has explored autonomous LLM agents that assist designers by automating script generation~\cite{openroad-agent-asu, openroadassistant}, parameter tuning, and flow configuration around EDA tools~\cite{ORFS-Agent}. OpenROAD Agent\cite{openroad-agent-asu} introduces a self-correcting LLM agent that generates Python-based OpenROAD scripts from natural language prompts, executes them within the tool environment, and iteratively refines outputs based on real-time error feedback.  ORFS-Agent\cite{ORFS-Agent}, operates as an iterative optimization agent within OpenROAD-flow-scripts (ORFS) that reads flow metrics, proposes new parameter configurations, and refines them through successive tool executions. Both these works use LLMs \textit{around} existing EDA tools.

\noindent
\textbf{Agentic Work for EDA.}
Two recent works have explored the use of agentic LLMs for source code modification or generation in EDA. \cite{AuDoPEDA} demonstrates the capability to operate over large EDA codebases by constructing structured representations, generating implementation plans grounded in domain knowledge, proposing code diffs, compiling modified tools, and evaluating QoR metrics such as wirelength, clock period, and power. \cite{yu-sat} shows that autonomous code evolution can be applied to NP-complete problems such as Boolean satisfiability, where multi-agent systems iteratively refine solver implementations under correctness and runtime constraints. Together, these works establish that LLMs can effectively generate and evolve repository-scale, performance-critical code. These directions have also been contextualized in prior work \cite{ispd26_review, date-asu-paper}, which distinguishes between flow-level automation and code-level agents that directly modify EDA tool internals.


\section{Design-adaptive Global Routing}\label{sec:design-adaptive-gr}

We introduce GR-Evolve, a design-adaptive code-evolution framework that transforms global routing from a fixed-algorithm problem into a program (source code or heuristic) optimization problem. Instead of applying a single static router across all designs, GR-Evolve automatically generates a specialized routing algorithm for each target design by iteratively modifying a baseline router source code and evaluating its QoR. At a high level, GR-Evolve operates as a closed-loop optimization system. Starting from a baseline global router, the framework repeatedly proposes code-level modifications, compiles and executes the modified router within the OpenROAD flow, and evaluates its performance. These evaluation results are fed back into the system to guide subsequent modifications, enabling progressive improvement over multiple iterations. As a result, GR-Evolve enables \emph{design–tool co-exploration}, where both the design and the underlying EDA tool evolve together.

\subsection{Problem Formulation}
\label{sec:problem-formulation}
\noindent
We formulate design-adaptive global routing as a \emph{program optimization problem}, where the objective is to automatically specialize a global routing algorithm to a given target design. Let $D$ denote a placed netlist for a target design, and let $R_0$ denote a baseline global router implementation (e.g., FastRoute~\cite{fastroute}, CUGR~\cite{CUGR}, or SPRoute~\cite{SPRoute}). Our goal is to generate a modified router $R^*$, derived from $R_0$ through a sequence of source-code transformations, such that routing QoR is improved over $R_0$ when evaluated on $D$. Formally, we seek:
\vspace{-3mm}
\[
R^* = \arg\min_{R \in \mathcal{P}(R_0)} \; \mathcal{L}(R, D),
\]
where $\mathcal{P}(R_0)$ denotes the space of valid program variants reachable from the baseline router through code modifications, and $\mathcal{L}(R, D)$ is a multi-objective cost function derived from routing QoR metrics.

\noindent
\textbf{Objective function.}
We treat $\mathcal{L}(R, D)$ as a multi-objective vector and approximate optimization via Pareto dominance under a prioritized objective hierarchy (DR WL > DR VC > GR RT). These objectives are inherently conflicting, requiring trade-offs rather than a single optimal solution. In practice, due to the objective's expensive, non-differentiable nature, we approximate this solution by bounded exploration of the program space.


\noindent
\textbf{Search space.}
The search space $\mathcal{P}(R_0)$ consists of syntactically valid modifications to the router source code, including changes to cost functions, routing heuristics, control flow, and data structures. This space is extremely large, spanning thousands of lines of code.

\noindent
\textbf{Key idea.}
Rather than optimizing routing solutions or tuning parameters, we search for the best-performing \emph{routing algorithm} by using an LLM as an agent to generate and search over program variants, enabling design-specific adaptation of routing heuristics.

\subsection{Research Challenges}
\label{sec:challenges}
Formulating global routing as a program optimization problem introduces several fundamental challenges that distinguish it from conventional design-space exploration and hyperparameter tuning.

\noindent
\textbf{(1) Large search space under limited context.}
The search space consists of program variants derived from modifying the source code of a global routing engine, which spans thousands of lines and complex interdependent modules. At the same time, modern chip designs involve large-scale data such as netlists, placement information, and routing resources. However, LLMs operate under strict context window limitations, making it infeasible to directly expose the full codebase and design state during optimization. This creates a fundamental tension between the scale of the search space and the limited contextual capacity available for reasoning.

\noindent
\textbf{(2) Expensive and delayed evaluation.}
Evaluating a single program variant requires executing global routing followed by detailed routing within a full EDA flow. This process can take minutes for small designs and hours for larger designs, making iterative optimization extremely costly. Furthermore, improvements introduced at the global routing stage may only manifest after detailed routing, resulting in delayed and sparse feedback signals. This significantly complicates the optimization process compared to settings with fast or immediate evaluation.

\noindent
\textbf{(3) Multi-objective optimization with conflicting metrics.}
Routing quality is inherently multi-objective, involving trade-offs between both DR and GR WL, VC and RT. Improvements in one objective may degrade others, and no single scalar metric fully captures routing quality. As a result, the optimization process must reason over trade-offs and explore a set of Pareto-optimal solutions rather than converging to a single optimum.

\noindent
These challenges require new mechanisms for context management, efficient exploration under expensive evaluation, and principled handling of multi-objective trade-offs, which are not addressed by existing approaches to global routing or EDA tool optimization.

\begin{figure}[t]
    \centering
    \includegraphics[width=0.42\textwidth]{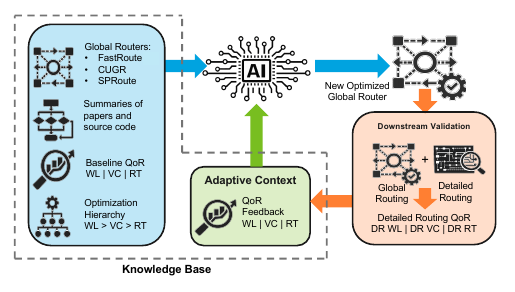}
    \vspace{-6mm}
    \caption{Overview of the GR-Evolve framework showing the closed-loop code evolution process, including knowledge base construction, context-driven code evolution, and downstream validation with QoR feedback.} 
    \label{fig:grevolve-overview}
    \vspace{-6mm}
\end{figure}

\subsection{GR-Evolve Framework}
\label{sec:gr-evolve}
To solve the problem outlined in Section~\ref{sec:problem-formulation} while addressing the challenges in Section~\ref{sec:challenges}, we propose GR-Evolve, a design-adaptive code-evolution framework that performs a bounded search over program variants, guided by QoR feedback, and maintains all evaluated candidates for post-hoc selection.
At a high level, the framework operates as a closed-loop system as shown in Fig.~\ref{fig:grevolve-overview}. Starting from a baseline global router, GR-Evolve repeatedly proposes code-level modifications, compiles and executes the modified router within the OpenROAD flow, and evaluates its QoR using post-detailed-routing metrics. The evaluation outcomes are fed back into the system to guide subsequent modifications, enabling progressive improvement over multiple iterations. As illustrated in Fig.~\ref{fig:grevolve-overview}, GR-Evolve consists of three key components: (1)~knowledge base for structured context construction, (2)~context-driven code evolution for iterative program optimization with implicit multi-objective reasoning, and (3)~downstream validation using detailed routing metrics. Together, these components enable program-level optimization of routing algorithms under expensive, multi-objective evaluation.

\begin{figure}[t]
    \centering
    \includegraphics[width=0.33\textwidth]{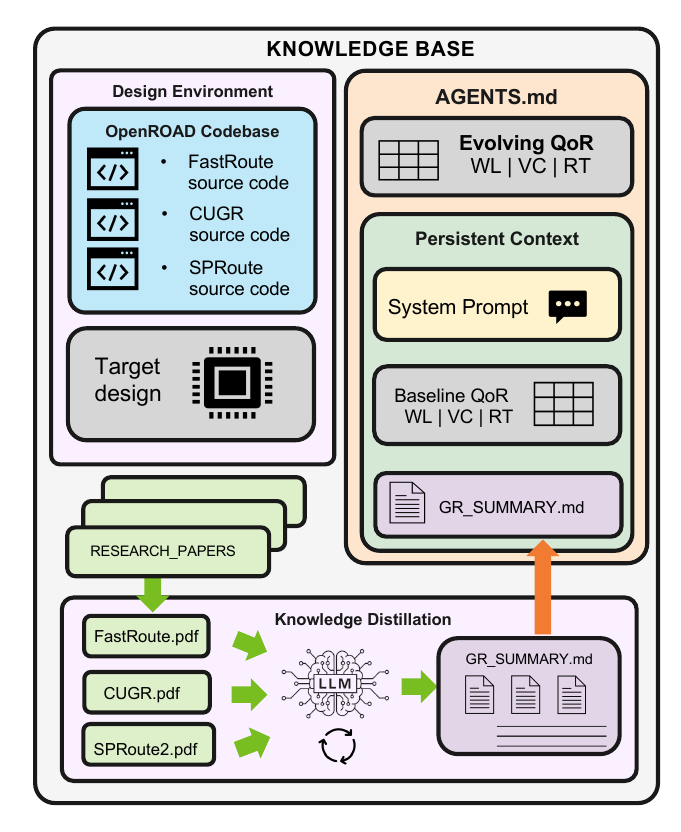}
    \vspace{-5mm}
    \caption{Knowledge base and context provided to GR-Evolve.}
    \vspace{-7mm}
    \label{fig:knowledge-distillation}
\end{figure}

\subsubsection{Knowledge Base}

\noindent
To address the challenge of large search space under limited context (Challenge 1 in Section~\ref{sec:challenges}), GR-Evolve constructs a structured knowledge base that provides the LLM agent with compact yet expressive domain information. The knowledge base consists of three components as shown in Fig.~\ref{fig:knowledge-distillation}.

\noindent
\textbf{(a) Design environment.}
This includes the OpenROAD codebase and OpenROAD-flow-scripts (ORFS) setup for the target design, including placed netlists, technology libraries, and routing configurations. This environment defines the execution context in which program variants are evaluated.

\noindent
\textbf{(b) Distilled algorithmic knowledge.}
Rather than exposing full research papers or entire codebases, we provide LLM-generated summaries of baseline routing algorithms (e.g., FastRoute~\cite{fastroute}, CUGR~\cite{CUGR}, SPRoute~\cite{SPRoute}) as shown in Fig.~\ref{fig:knowledge-distillation}. These summaries capture key design principles such as cost formulations, routing strategies, and data structures, enabling high-level reasoning within limited context. Effective agentic reasoning requires structured domain context alongside execution feedback~\cite{react-test},  hence we provide the agent with complementary forms of knowledge: source code access for implementation-level reasoning, and distilled algorithm summaries for design-intent comprehension. This ensures the agent can reason about \textit{what} to change, informed by algorithmic understanding, and \textit{how} to change it, informed by access to the code.

\noindent
\textbf{(c) Persistent context.} This consists of three key elements that guide the agent throughout the evolution process. First, it includes the baseline QoR metrics for each global router, which establish a reference point for evaluating improvements. Second, it incorporates the distilled summaries of global routing algorithms, providing high-level algorithmic understanding.

Third, the persistent context includes a system prompt that defines the agent’s operating procedure for the entire evolution. The system prompt contains three main components: 
(1) the optimization objectives and their priority hierarchy, explicitly encoding primary, secondary, and tertiary goals; 
(2) the execution pipeline, which structures each iteration into a repeatable cycle of reading context, planning a modification, implementing the change, compiling the modified router, executing it on the target design, and recording the resulting metrics; and 
(3) the read/write scope, which restricts the agent’s modifications to the evolved router directory while granting read access to the broader OpenROAD codebase.

This scope restriction serves as a safety mechanism that preserves the integrity of the reference router implementations, ensuring that baseline comparisons remain valid throughout the evolution.   Together, these components compress domain knowledge into a form that can be efficiently consumed by the LLM, enabling informed program modifications despite context limitations.

\subsubsection{Context-Driven Code Evolution}

\noindent
To address both the large search space and expensive evaluation (Challenges 1 and 2 of Section~\ref{sec:challenges}), GR-Evolve performs iterative program optimization through context-driven code evolution. At each iteration, the agent constructs a context consisting of the persistent knowledge base and the accumulated QoR history from prior iterations. Based on this context, the agent proposes a code-level modification (mutation) to the global routing algorithm. The modified router is then compiled and executed within the OpenROAD flow (only global routing and detailed routing), and its QoR metrics are evaluated. The resulting metrics are appended to a persistent history table, which records the evolution trajectory across iterations and serves as the primary feedback signal for subsequent decisions.

\noindent
\textbf{Stateless architecture.}
To address the challenges of limited context and long-horizon optimization (Challenges 1 and 2), GR-Evolve adopts a stateless execution model in which all state is externalized into persistent artifacts rather than maintained through conversational memory where each iteration is executed as a fresh agentic session in headless mode. At the start of every iteration, the agent reconstructs its working context by reading the QoR history, version control logs, and the current router source code. This design ensures that context remains bounded and consistent across long optimization runs, avoiding degradation in reasoning quality associated with extended interaction histories~\cite {LongContext}, and provides the agent with the freedom to select its context. 

The agent operates on a dedicated copy of the base router source code, where all modifications accumulate in-place. Each mutation is recorded as a Git commit, capturing both the source-code changes and the corresponding QoR metrics. This produces a complete version-controlled history of the evolution process. To ensure safety and reproducibility, the agent is restricted to write access within this sandboxed directory while retaining read access to the broader OpenROAD codebase. This restriction preserves the integrity of baseline router implementations and ensures that all comparisons remain valid.  Within each iteration, the agent follows a structured pipeline: it reviews the QoR history and prior modifications, reasons about potential improvements, proposes a code change, implements the modification, compiles the updated router, executes it on the target design, and records the resulting metrics. This stateless design is particularly important given the high cost of evaluation in EDA workflows, where a single iteration may take minutes to hours and full evolution runs may span multiple days. By reconstructing context from persistent artifacts at each iteration, we enable stable long-horizon optimization without exceeding context limits.

\noindent
\textbf{Self-repair mechanism.}
If a proposed modification introduces compilation or runtime errors, the agent is allowed to diagnose and repair the issue within the same iteration, enabling uninterrupted optimization without manual intervention.

\noindent
\textbf{Warm-start evolution.}
To mitigate the high cost of evaluation (Challenge 2), GR-Evolve supports warm-start evolution, where optimized routers on smaller designs are used to initialize evolution for larger designs. This reduces the number of required evolution iterations by transferring learned heuristics across designs.

\noindent
\textbf{Multi-objective reasoning.}
To address the multi-objective nature of routing optimization (Challenge 3), the agent is guided by an input optimization hierarchy while simultaneously maintaining a comprehensive log of all evaluated program variants. The optimization hierarchy, encoded in the system prompt, provides a soft prioritization among objectives (e.g., prioritizing DR WL over DR VC and GR RT), guiding the agent’s decision-making during each iteration. The framework records QoR metrics for each mutation across iterations, thereby constructing an explicit trade-off space among competing objectives. Each generated mutation is version-controlled and Git-logged, capturing both the source-code changes and the associated QoR metrics. 

By retaining all evaluated variants, the framework decouples exploration from selection, allowing the agent to explore diverse routing strategies, including intermediate changes that may temporarily degrade QoR. After evolution concludes, the QoR metrics table is used to identify the best-performing router, and the corresponding commit is checked out. This process is a manual post-hoc selection of solutions that best satisfy desired QoR trade-offs, enabling post-hoc identification of non-dominated solutions.

\subsubsection{Downstream Validation}
\noindent
Each iteration executes both GR and DR under iso-placement conditions, where placement is fixed across all runs. This ensures that differences in QoR are attributable solely to the routing algorithm. We track three metrics: DR WL, DR VC and GR RT. By using DR metrics as the primary feedback signal, the framework avoids optimizing surrogate objectives that may not correlate with final routing quality. The complete QoR history across iterations serves both as feedback for the agent and as the selection mechanism for identifying the best router.

\begin{table}[h]
\centering
\caption{Benchmark suite used to evaluate GR-Evolve.}
\vspace{-2mm}
\label{tab:design-nets}
\scriptsize
\setlength{\tabcolsep}{3pt}
\begin{tabular}{l|c|c|r|r}
\toprule
\textbf{Design} & \textbf{Design Nickname} & \textbf{Tech Node} & \textbf{\#Nets} & \textbf{Util.} \\
\midrule

\multirow{3}{*}{AES} 
 & \multirow{3}{*}{AES} & \texttt{SKY130HD}    & 15909  & 20\% \\
 &                     & \texttt{Nangate45} & 18184  & 40\% \\
 &                     & \texttt{ASAP7}     & 18252  & 40\% \\

\midrule

\multirow{3}{*}{IBEX} 
 & \multirow{3}{*}{IBX} & \texttt{SKY130HD}    & 15676  & 45\% \\
 &                     & \texttt{Nangate45} & 18199  & 50\% \\
 &                     & \texttt{ASAP7}     & 21269  & 40\% \\

\midrule

\multirow{3}{*}{JPEG} 
 & \multirow{3}{*}{JPG} & \texttt{SKY130HD}    & 49169  & 50\% \\
 &                     & \texttt{Nangate45} & 76488  & 80\% \\
 &                     & \texttt{ASAP7}     & 64378  & 70\% \\

\midrule

\multirow{2}{*}{SWERV} 
 & \multirow{2}{*}{SWV} & \texttt{Nangate45} & 102474 & 40\% \\
 &                     & \texttt{ASAP7}     & 128615 & 40\% \\

\midrule

\multirow{2}{*}{DYNAMIC\_NODE} 
 & \multirow{2}{*}{DYN} & \texttt{Nangate45} & 14093  & 40\% \\
 &                     & \texttt{ASAP7}     & 17305  & 40\% \\

\midrule

\multirow{1}{*}{ARIANE136} 
 & AR136 & \texttt{Nangate45} & 203041 & 36\% \\

\midrule

\multirow{1}{*}{BLACK\_PARROT} 
 & BP    & \texttt{Nangate45} & 278614 & 40\% \\

\bottomrule
\end{tabular}
\vspace{-5mm}
\end{table}

\begin{figure*}[t]
    \centering
    \includegraphics[width=0.9\linewidth]{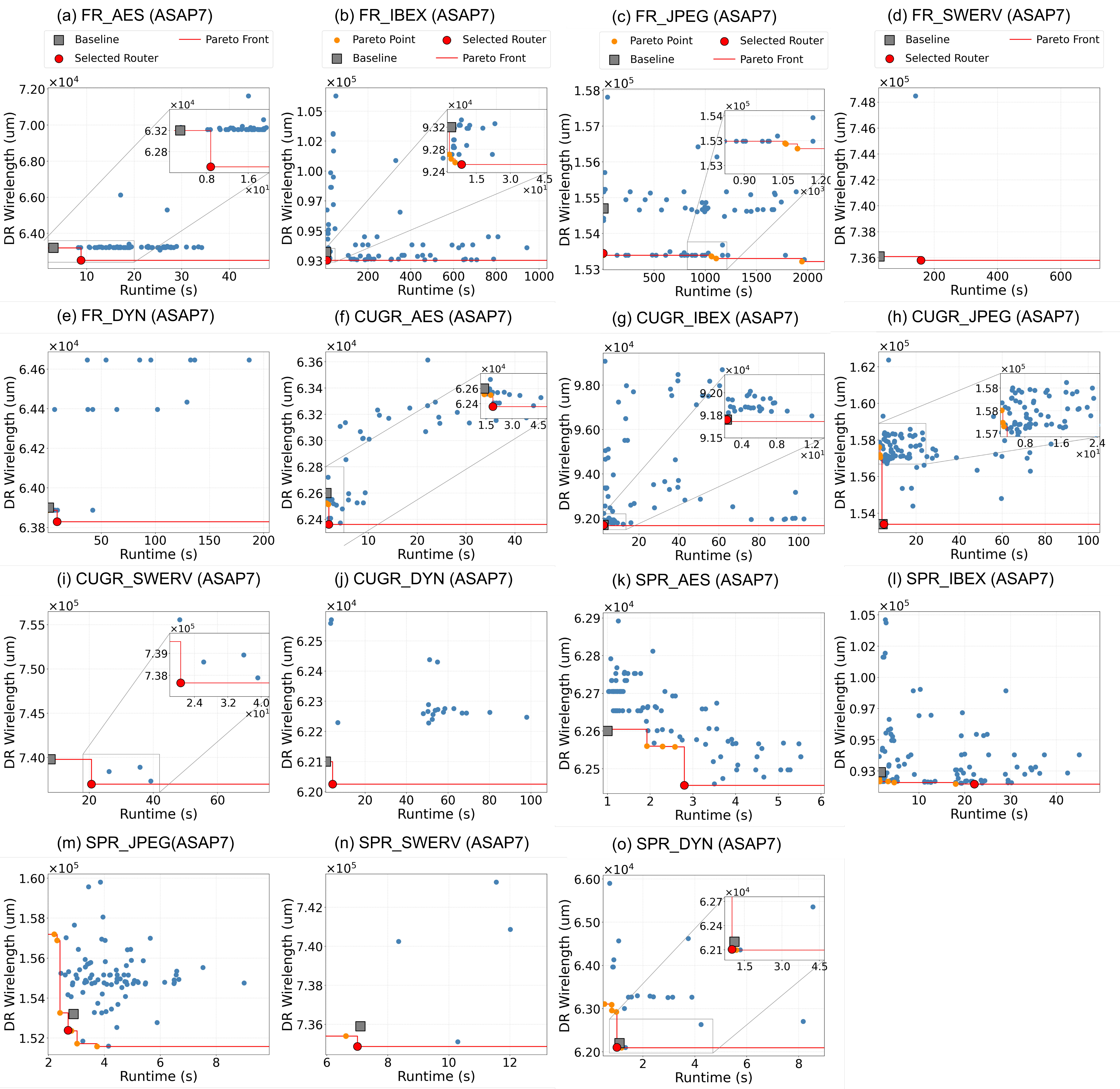}
    \vspace{-5mm}
    \caption{Pareto fronts of the search space of all 15 router-design pairs in ASAP7.  Selected router QoR is reported in \Cref{tab:asap7_dr_wl_gr_rt}. }
    \label{fig:pareto-front-plot-asap7}
    \vspace{-5mm}
\end{figure*}

\section{Experimental Setup}\label{subsec:experimental-setup}

\noindent
\textbf{Benchmarks and Technology Nodes.} To evaluate the effectiveness of GR-Evolve, we conduct experiments across three technology nodes: the SkyWater 130\,nm (SKY130HD)~\cite{skywater130-pdk}, the NanGate 45\,nm (NanGate45)~\cite{Nangate45}, and the ASAP7 7\,nm~\cite{asap7}  technology node. The designs and their corresponding details are given in Table~\ref{tab:design-nets}. All designs are synthesized and placed using the OpenROAD-flow-scripts (ORFS)~\cite{OpenROAD-flow-scripts} infrastructure with default settings, including default utilization targets and flow parameters.

In the NanGate45 technology node, we additionally include two larger-scale designs, BlackParrot and Ariane136, enabled through ORFS. For these designs, we employ the warm-start strategy described in Section~\ref{sec:gr-evolve}, where routing evolution is initialized from previously evolved router variants to improve convergence and evaluation efficiency on larger instances.  In addition to the standard design suite, we evaluate our approach on benchmark circuits from the ICCAD 2019 global routing contest~\cite{iccad19-contest}.

\noindent
\textbf{Evolution Setup.}
We evolve routers starting from three baseline routers: FastRoute (FR), CUGR, and SPRoute (SPR), producing 21 evolved routers on NanGate45 (7 designs), 15 routers on ASAP7 (5 designs), and 9 routers on SKY130HD (3 designs). Each evolved router is named by its base router and the design it is being evolved for; for example, \texttt{FR\_AES} denotes the variant evolved from FastRoute on AES.  For each design and baseline router, we run an iterative evolution process with up to 75 iterations. For larger designs, we run 25 iterations after a warm start.  

In each iteration, the agent proposes code-level modifications to the router, which are then compiled and evaluated within the ORFS flow. Evolution runs for a fixed number of iterations, after which the best-performing router from the explored set is selected. Each mutation corresponds to a program variant in the search space $\mathcal{P}(R_0)$ and is version-controlled and Git logged along with its QoR metrics, enabling reconstruction of the full optimization trajectory. 

\noindent
\textbf{Evaluation Methodology and Metrics.}
In the results, we report WL, VC, and RT across both GR and DR. Since the detailed router refines the global routing solution to produce the final physical wiring, DR metrics represent most direct measure of final routing quality. For the optimization hierarchy defined as input in section~\ref{sec:gr-evolve}, we ask the agent to prioritize DR WL, then DR VC, and finally GR RT.  All comparisons are made relative to the respective base router's unmodified baseline. 
All experiments are conducted under iso-placement conditions, where placement is fixed across all router variants. This ensures that differences in QoR are attributable solely to changes in the routing algorithm, consistent with the program-level optimization objective.

\noindent
\textbf{DRC and Routing Solution Correctness.} Routing feasibility is treated as a hard constraint rather than an optimization objective. Accordingly, while design rule check (DRC) violations are critical for ensuring valid routing solutions, we do not use DRC count as an optimization metric. We check DRC only for the final version of each selected router. Once a best router is identified, we run it on the design to generate metrics and check the DRC logs. In these final evaluations, all selected evolved routers produce DRC-clean solutions under detailed routing in the default ORFS setup. As a result, DRC provides insufficient variation to guide optimization. Instead, we focus on optimizing DR WL and VC once feasibility has been established.
All final designs are verified for correctness using layout-versus-schematic (LVS) checks. LVS is not performed at every evolution iteration because the netlist remains unchanged, and any inconsistencies generated by the LLM in the source code would be detected as mismatches between the netlist and the generated global routes in the OpenROAD database. 

\noindent
\textbf{Implementation and Runtime Environment.}
All experiments are conducted on a dual-socket AMD EPYC 7313 server with 32 cores, 64 threads, and 512 GB of RAM. Each iteration involves compiling the modified router and executing both global and detailed routing, resulting in runtimes ranging from minutes for smaller designs to up to an hour for larger designs. Full evolution runs span multiple iterations and may take several hours to days. All experiments use OpenAI Codex-5.3 with medium reasoning effort. The GR-Evolve framework is model-agnostic and can be adapted to other agents or LLMs without modification.

\noindent
\textbf{Evaluation.}
We evaluate the effectiveness of GR-Evolve in exploring the program space under the multi-objective formulation described in Section~\ref{sec:gr-evolve}. First, we evaluate the agent's ability to search the multi-objective space and generate multiple router variants. Second, we measure the quality of the selected routers in terms of WL, VC, and RT at both the GR and DR stages.

\subsection{Design-tool Co-exploration}

\noindent
We apply GR-Evolve to all design-router pairs across the three technology nodes (7 designs $\times$ 3 routers in NanGate45, 5 designs $\times$ 3 routers each in ASAP7, and 3 designs $\times$ 3 routers in SKY130HD), shown in Table~\ref{tab:design-nets}. We present the Pareto fronts for all iterations of the 15 design--router pairs in ASAP7 in  Fig.~\ref{fig:pareto-front-plot-asap7}. The
density of points varies across subplots: router–design pairs such
as FR\_JPEG, SPR\_JPEG, and CUGR\_IBEX exhibit dense exploration
with many evaluated mutations, reflecting evolution runs that proceeded through a larger number of iterations. In contrast, sparser point clouds correspond to runs that had several failures during evolutions due to errors during detailed routing, or DRC violations.   Across the majority of subplots,
the baseline router (gray square) lies above the
Pareto front, indicating that GR-Evolve discovers router variants
that improve over the baseline in DR WL, GR RT, or both. In some cases, the improvements are minor compared to the gray baseline.

Example, in  Fig.~\ref{fig:pareto-front-plot-asap7} (f) CUGR\_AES and (k) SPR\_AES, multiple non-dominated variants are available along the frontier, enabling WL–RT trade-off exploration without re-running the evolution process. This decoupling of exploration from selection is central to the design–tool co-exploration paradigm introduced in \Cref{sec:design-adaptive-gr}.

\begin{figure*}[t]
    \centering
    \includegraphics[width=0.9\linewidth]{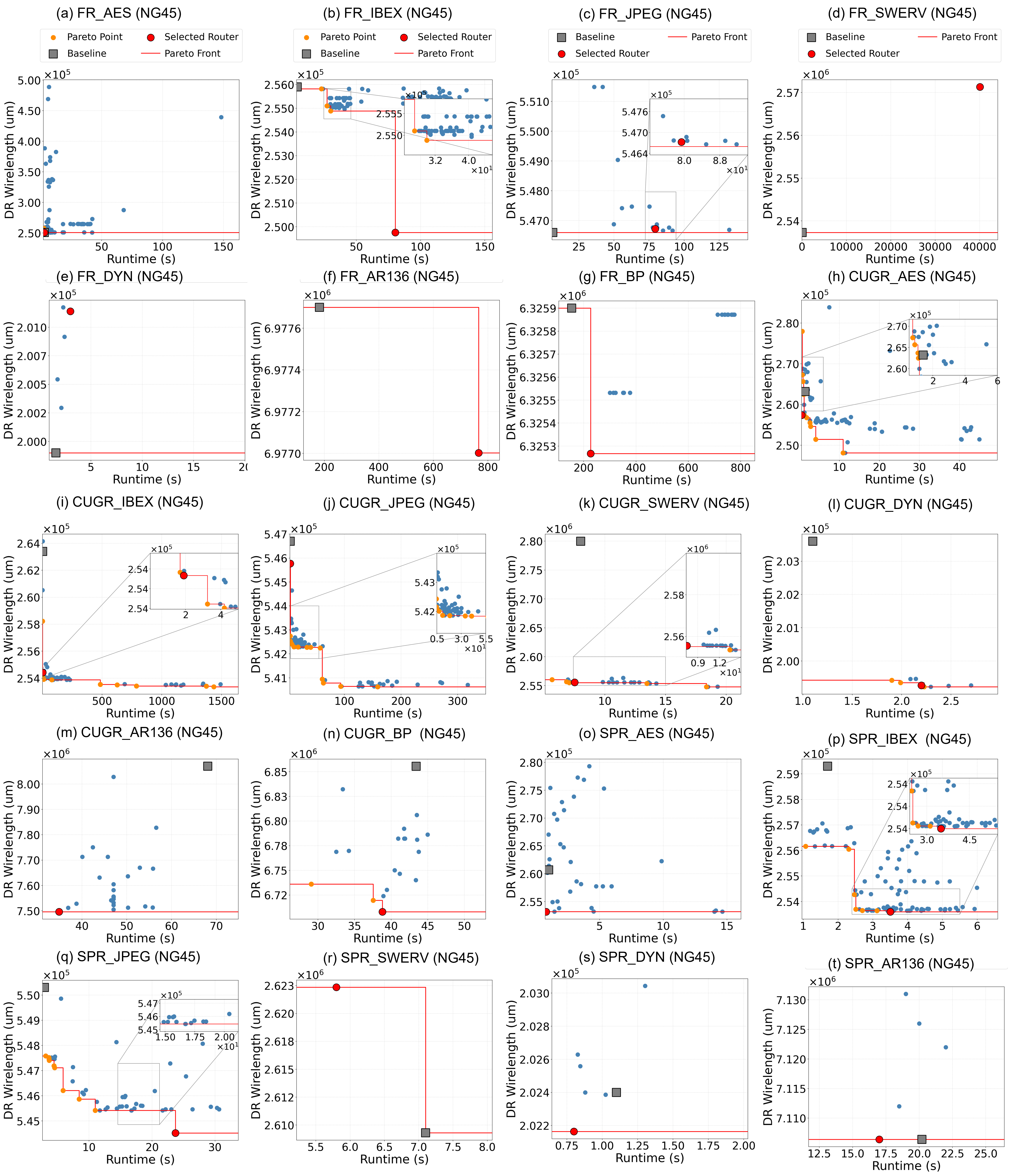}
    \vspace{-5mm}
    \caption{Pareto fronts of the search space for 20 of the 21 router-design pairs in Nangate45 (NG45); SPR\_BP is omitted (but can be found on our repository~\cite{gr-evolve-gh}). QoR metrics of the selected router for all 21 pairs are reported in \Cref{tab:nangate45_dr_wl_gr_rt}.}
    \label{fig:pareto-front-plot-nangate45}
      \vspace{-6mm}
\end{figure*}

Fig.~\ref{fig:pareto-front-plot-nangate45} presents the Pareto fronts for 20 of the 21 router–design pairs in the Nangate45 technology node, covering FastRoute, CUGR, and SPRoute baselines across AES, IBEX, JPEG, SWERV, DYN, AR136, and BP. As in the ASAP7 results, the baseline router lies off the Pareto frontier in most subplots, confirming that the evolved routers achieve improvements over their corresponding baselines across a diverse range of design sizes and routing characteristics. In a smaller number of cases, such as Fig.\ref{fig:pareto-front-plot-nangate45}(d) FR\_SWERV, (e) FR\_DYN, and (r) SPR\_SWERV, the baseline lies on or close to the Pareto front, indicating that the explored variants either match the baseline or offer only marginal improvements for those particular design–router pairs. Subplots corresponding to smaller designs, such as Fig.~\ref{fig:pareto-front-plot-nangate45}(a) FR\_AES, (h) CUGR\_AES, and (p) SPR\_IBEX, exhibit dense exploration with a large number of evaluated mutations, reflecting the full 75-iteration evolution budget used for these smaller designs. In contrast, subplots corresponding to warm-started larger designs such as Fig.~\ref{fig:pareto-front-plot-nangate45}(f) FR\_AR136, (g) FR\_BP, (m) CUGR\_AR136, (n) CUGR\_BP and (t) SPR\_AR136 exhibit noticeably fewer points, since these runs use only 25 or fewer iterations initialized from previously evolved variants, as described in \Cref{subsec:experimental-setup}.

\begin{figure*}[t]
    \centering
    \includegraphics[width=0.9\linewidth]{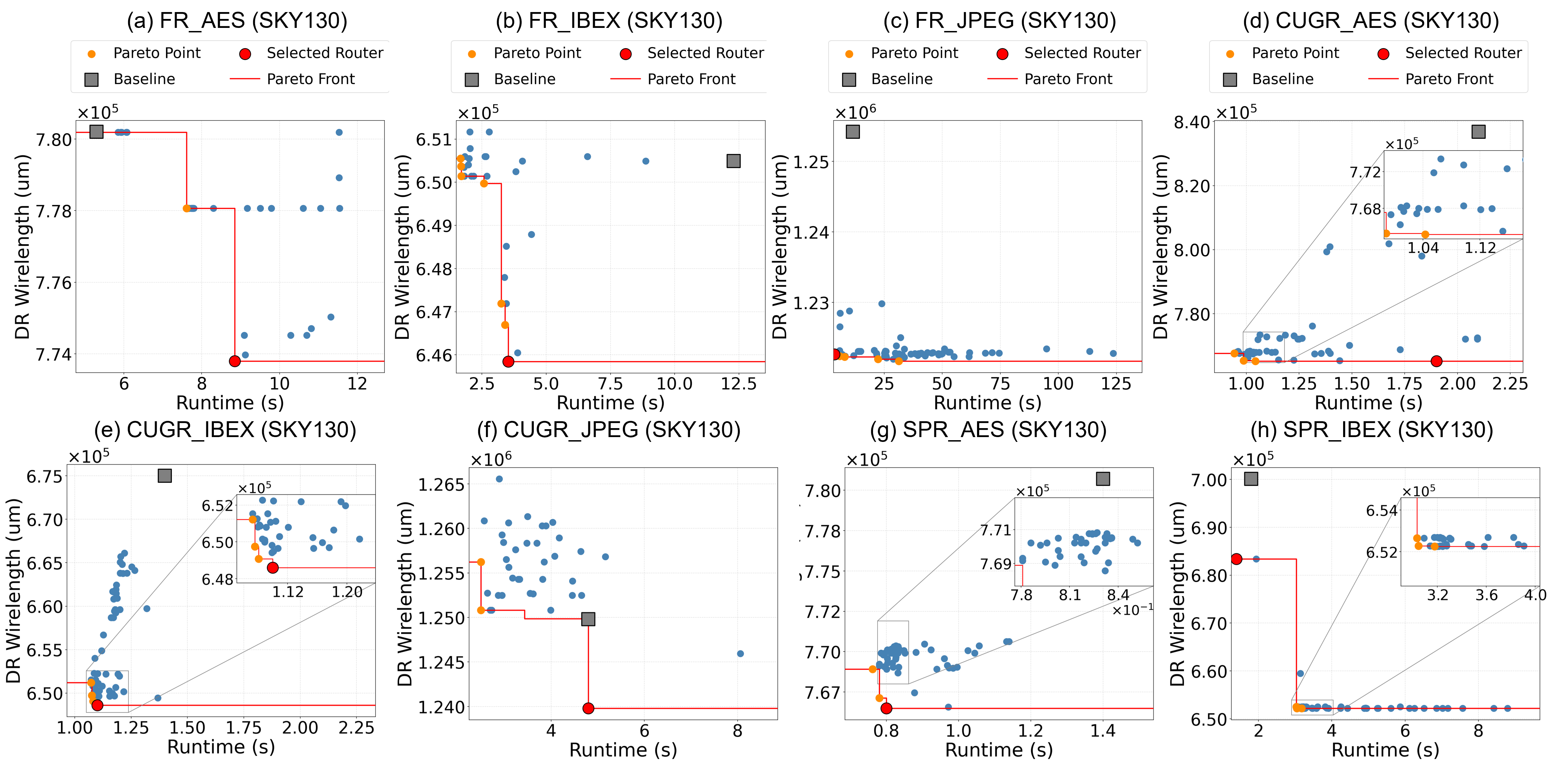}
    \vspace{-5mm}
    \caption{Pareto fronts of the evolution search space for 8 of the 9 router-design pairs in SKY130HD; SPR\_JPEG is omitted (but can be found on our repository~\cite{gr-evolve-gh}). QoR metrics for the selected routers are reported in \Cref{tab:sky130_dr_wl_gr_rt}.}
    \label{fig:pareto-front-plot-sky130}
    \vspace{-5mm}
\end{figure*}

Fig.~\ref{fig:pareto-front-plot-sky130} presents the Pareto fronts for 8 of the 9 router–design pairs in the SKY130HD technology node. The baseline router lies above or to the right of the Pareto frontier in nearly all cases, demonstrating that the improvements observed from GR-Evolve generalize to SKY130HD across all three baseline routers. Variants such as \Cref{fig:pareto-front-plot-sky130}(d) CUGR\_AES show substantial DR WL reductions over their corresponding baselines.

\subsection{QoR of Selected Routers}\label{subsec:qor-selection}
From the search space illustrated in Fig.~\ref{fig:pareto-front-plot-asap7},~\ref{fig:pareto-front-plot-nangate45}, and ~\ref{fig:pareto-front-plot-sky130} we select, for each design--router pair, a single router that achieves improvement in both post-DR WL and GR runtime than the baseline router. If no such router exists, we instead select the router with the best post-DR WL irrespective of runtime and evaluate its QoR against the corresponding baseline routers. In designs where no variant Pareto-dominates the baseline in both DR WL and GR RT, the fallback selection (best DR WL) is applied; this explains the runtime degradations observed in some rows of Tables~\ref{tab:asap7_dr_wl_gr_rt},~\ref{tab:nangate45_dr_wl_gr_rt}, and~\ref{tab:sky130_dr_wl_gr_rt}. This selected router is highlighted by the red point in the figures.
Tables~\ref{tab:asap7_dr_wl_gr_rt},~\ref{tab:nangate45_dr_wl_gr_rt}, and~\ref{tab:sky130_dr_wl_gr_rt} present the QoR for the selected routers across the ASAP7, Nangate45, and SKY130HD, respectively.

For each design–router pair, we report both GR and DR metrics, including WL and VC, and total WL (TWL, which is the sum of WL and VC). For the baseline routers, we report the absolute values of GR and DR WL and VC, providing a direct reference for the routing quality achieved by the unmodified tools. DR RT is reported for completeness but is not an optimization objective; only GR RT is used in the optimization hierarchy. In contrast, for the evolved routers, we report percentage improvements relative to their corresponding baselines. We also report the number of lines of code modified (LoCM) for each evolved router.

\begin{table}[t]
\centering
\caption{QoR Metrics post GR and post DR of selected evolved router for ASAP7. Positive $\Delta$WL denotes reduction.}
\label{tab:asap7_dr_wl_gr_rt}
\scriptsize
\setlength{\tabcolsep}{2.5pt}
\resizebox{\columnwidth}{!}{%
\begin{tabular}{l|l|cccc|cccc|c}
\toprule
 & & \multicolumn{4}{c|}{\textbf{GR}} & \multicolumn{4}{c|}{\textbf{DR}} & \\
 & & WL($\mu$m) & VC & TWL($\mu$m) & & WL($\mu$m) & VC & TWL($\mu$m) & & \\
\textbf{Des.} & \textbf{Router} & $\Delta$WL & $\Delta$VC & $\Delta$TWL & RT(s) & $\Delta$WL & $\Delta$VC & $\Delta$TWL & RT(s) & \textbf{\shortstack{\#LoCM}}\\
\midrule
\multirow{6}{*}{AES}
 & FastRoute   & 88.0k & 138.1k & 226.1k & 3.0    & 63.2k & 160.3k & 223.5k & 612.4 & -- \\
 & CUGR        & 85.5k & 132.4k & 217.9k & 1.4    & 62.6k & 160.4k & 222.9k & 667.9 & -- \\
 & SPRoute     & 87.4k & 112.7k & 200.1k & 1.0    & 62.6k & 160.1k & 222.7k & 621.1 & -- \\
 & FR\_AES     & \textcolor{green!70!black}{+2.86} & \textcolor{green!70!black}{+4.13} & \textcolor{green!70!black}{+3.63} & 8.8 & \textcolor{green!70!black}{+1.09} & \textcolor{red}{$-$0.07} & \textcolor{green!70!black}{+0.26} & \textcolor{green!70!black}{600.3 }& 624 \\
 & CUGR\_AES   & \textcolor{green!70!black}{+0.11} & \textcolor{red}{$-$5.38} & \textcolor{red}{$-$3.22} & 1.9  & \textcolor{green!70!black}{+0.38} & \textcolor{red}{$-$1.74} & \textcolor{red}{$-$1.14} & \textcolor{green!70!black}{402.1 }& 1107 \\
 & SPR\_AES    & \textcolor{green!70!black}{+3.27} & \textcolor{red}{$-$23.81} & \textcolor{red}{$-$11.98} & 2.8  & \textcolor{green!70!black}{+0.23} & \textcolor{red}{$-$0.09} & \textcolor{green!70!black}{+0.00} & \textcolor{green!70!black}{569.3}& 1965 \\
\midrule
\multirow{6}{*}{IBX}
 & FastRoute   & 124.6k & 173.6k & 298.3k & 3.9    & 93.2k & 205.2k & 298.4k & 684.8  & -- \\
 & CUGR        & 121.6k & 169.0k & 290.7k & 2.3    & 91.7k & 201.4k & 293.1k & 781.0  & -- \\
 & SPRoute     & 123.5k & 134.1k & 257.7k & 1.7    & 92.4k & 204.7k & 297.1k & 786.7  & -- \\
 & FR\_IBX    & \textcolor{red}{$-$1.00} & \textcolor{red}{$-$10.19} & \textcolor{red}{$-$6.35} & 8.5  & \textcolor{green!70!black}{+0.72} & \textcolor{green!70!black}{+3.45} & \textcolor{green!70!black}{+2.60} & 825.8 &1212 \\
 & CUGR\_IBX  & 0.00 & 0.00 & 0.00 & 2.3  & 0.00 & 0.00 & 0.00 & 1004.9 & 90 \\
 & SPR\_IBX   & \textcolor{green!70!black}{+7.78} & \textcolor{red}{$-$1.92} & \textcolor{green!70!black}{+2.73} & 22.1 & \textcolor{green!70!black}{+1.02} & \textcolor{green!70!black}{+0.83} & \textcolor{green!70!black}{+0.89} & 905.9 &5762 \\
\midrule
\multirow{6}{*}{JPG}
 & FastRoute   & 230.8k & 429.4k & 660.2k & 6.4      & 154.7k & 459.4k & 614.0k & 1130.1 & -- \\
 & CUGR        & 221.7k & 414.6k & 636.3k & 4.4      & 153.4k & 462.4k & 615.8k & 1179.0 & -- \\
 & SPRoute     & 229.0k & 320.2k & 549.2k & 2.9      & 153.2k & 457.5k & 610.7k & 1048.0 & -- \\
 & FR\_JPG    & \textcolor{green!70!black}{+3.91} & \textcolor{green!70!black}{+3.46} & \textcolor{green!70!black}{+3.62} & \textcolor{green!70!black}{2.7} & \textcolor{green!70!black}{+0.81} & \textcolor{red}{$-$0.70} & \textcolor{red}{$-$0.32} & \textcolor{green!70!black}{413.3  }&  699\\
 & CUGR\_JPG  & 0.00 & 0.00 & 0.00 & 5.0    & 0.00 & 0.00 & 0.00 & \textcolor{green!70!black}{1119.6 }& 90 \\
 & SPR\_JPG   & \textcolor{red}{$-$9.32} & \textcolor{red}{$-$20.01} & \textcolor{red}{$-15.55$} & \textcolor{green!70!black}{2.7}    & \textcolor{green!70!black}{+0.53} & \textcolor{red}{$-$1.83} & \textcolor{red}{$-$1.24} & \textcolor{green!70!black}{542.9 }&  1472\\
\midrule
\multirow{6}{*}{SWV}
 & FastRoute   & 920.0k & 1055.2k & 1975.3k & 27.3  & 736.1k & 1286.2k & 2022.3k & 1445.6 & -- \\
 & CUGR        & 912.9k & 1038.6k & 1951.5k & 8.0   & 739.8k & 1289.8k & 2029.6k & 1289.8 & -- \\
 & SPRoute     & 922.3k & 835.9k & 1758.2k & 7.1   & 735.9k & 1299.1k & 2035.0k & 1374.8 & -- \\
 & FR\_SWV   & \textcolor{green!70!black}{+0.02} & 0.00 & \textcolor{green!70!black}{+0.01} & 157.7 & \textcolor{green!70!black}{+0.04} & \textcolor{green!70!black}{+0.08} & \textcolor{green!70!black}{+0.07} & 1795.0 & 451 \\
 & CUGR\_SWV & \textcolor{green!70!black}{+0.69} & \textcolor{green!70!black}{+5.30} & \textcolor{green!70!black}{+3.14} & 20.7  & \textcolor{green!70!black}{+0.38} & \textcolor{red}{$-$1.82} & \textcolor{red}{$-$1.02} & 1387.8 & 599 \\
 & SPR\_SWV  & \textcolor{red}{$-$0.22} & \textcolor{red}{$-$2.25} & \textcolor{red}{$-$1.18} & \textcolor{green!70!black}{7.0}   & \textcolor{green!70!black}{+0.14} & \textcolor{red}{$-$0.57} & \textcolor{red}{$-$0.31} & 1424.4 & 1136 \\
\midrule
\multirow{6}{*}{DYN}
 & FastRoute   & 85.4k & 122.0k & 207.4k & 1.6 & 63.9k & 142.9k & 206.8k & 374.0 & -- \\
 & CUGR        & 82.6k & 116.6k & 199.2k & 1.1 & 62.1k & 139.1k & 201.2k & 291.0 & -- \\
 & SPRoute     & 84.1k & 92.2k & 176.3k & 1.1 & 62.2k & 139.4k & 201.6k & 432.1 & -- \\
 & FR\_DYN      & \textcolor{green!70!black}{+3.33} & \textcolor{green!70!black}{+29.69} & \textcolor{green!70!black}{+18.83} & 9.2 & \textcolor{green!70!black}{+0.11} & \textcolor{green!70!black}{+0.10} & \textcolor{green!70!black}{+0.11} & \textcolor{green!70!black}{260.6} & 451 \\
 & CUGR\_DYN    & \textcolor{green!70!black}{+0.02} & \textcolor{red}{$-$3.74} & \textcolor{red}{$-$2.18} & 4.3 & \textcolor{green!70!black}{+0.12} & \textcolor{red}{$-$0.72} & \textcolor{red}{$-$0.46} & \textcolor{green!70!black}{223.8} & 618 \\
 & SPR\_DYN     & \textcolor{red}{$-$1.70} & \textcolor{red}{$-$25.95} & \textcolor{red}{$-$14.38} & \textcolor{green!70!black}{1.0} & \textcolor{green!70!black}{+0.15} & \textcolor{red}{$-$2.86} & \textcolor{red}{$-$1.93} & \textcolor{green!70!black}{216.8} & 961 \\
\bottomrule
\end{tabular}
}%
\end{table}

\begin{table}[t]
\centering
\caption{QoR Metrics post GR and post DR of selected evolved router for Nangate45. Positive $\Delta$WL denotes a reduction.}
\vspace{-2mm}
\label{tab:nangate45_dr_wl_gr_rt}
\scriptsize
\setlength{\tabcolsep}{2.5pt}
\resizebox{\columnwidth}{!}{%
\begin{tabular}{l|l|cccc|cccc|c}
\toprule
 & & \multicolumn{4}{c|}{\textbf{GR}} & \multicolumn{4}{c|}{\textbf{DR}} & \\
 & & WL($\mu$m) & VC & TWL($\mu$m) & & WL($\mu$m) & VC & TWL($\mu$m) & & \\
\textbf{Design} & \textbf{Router} & $\Delta$WL & $\Delta$VC & $\Delta$TWL & RT(s) & $\Delta$WL & $\Delta$VC & $\Delta$TWL & RT(s) & \textbf{\shortstack{\#LoCM}}\\
\midrule
\multirow{6}{*}{AES}
 & FastRoute   & 352.4k & 160.9k & 513.3k & 3.0    & 250.9k & 136.7k & 387.7k & 612.4 & -- \\
 & CUGR        & 363.5k & 167.6k & 531.1k & 1.4    & 263.2k & 146.9k & 410.1k & 667.9 & -- \\
 & SPRoute     & 366.8k & 133.2k & 500.0k & 1.0    & 260.7k & 140.8k & 401.5k & 621.1 & -- \\
 & FR\_AES     & \textcolor{green!70!black}{+0.13} & \textcolor{green!70!black}{+0.03} & \textcolor{green!70!black}{+0.10}   & 3.0  & \textcolor{green!70!black}{+0.04} & \textcolor{red}{$-$0.03} & \textcolor{green!70!black}{+0.02} & \textcolor{green!70!black}{190.2 } & 561 \\
 & CUGR\_AES   & \textcolor{green!70!black}{+1.13} & \textcolor{green!70!black}{+31.35} & \textcolor{green!70!black}{+10.67} & \textcolor{green!70!black}{0.6} & \textcolor{green!70!black}{+2.21} & \textcolor{green!70!black}{+1.55} & \textcolor{green!70!black}{+1.97} & \textcolor{green!70!black}{373.0} &  5341\\
 & SPR\_AES    & \textcolor{green!70!black}{+7.72} & \textcolor{green!70!black}{+0.27} & \textcolor{green!70!black}{+5.74}   & \textcolor{green!70!black}{0.8}  & \textcolor{green!70!black}{+2.89} & \textcolor{red}{$-$0.15} & \textcolor{green!70!black}{+1.82} & \textcolor{green!70!black}{348.9} & 1977 \\
\midrule
\multirow{6}{*}{IBX}
 & FastRoute   & 355.4k & 147.4k & 502.8k & 3.9    & 255.9k & 126.7k & 382.6k & 684.8  & -- \\
 & CUGR        & 357.4k & 156.8k & 514.2k & 2.3    & 263.4k & 138.3k & 401.7k & 781.0  & -- \\
 & SPRoute     & 361.5k & 118.1k & 479.6k & 1.7    & 259.3k & 130.7k & 389.9k & 786.7  & -- \\
 & FR\_IBX    & \textcolor{green!70!black}{+2.29} & \textcolor{red}{$-$9.22} & \textcolor{red}{$-$1.08}                    & 80.3 & \textcolor{green!70!black}{+2.40} & \textcolor{red}{$-$8.73} & \textcolor{red}{$-$1.23} & \textcolor{green!70!black}{147.6} & 745 \\
 & CUGR\_IBX  & \textcolor{green!70!black}{+4.26} & \textcolor{green!70!black}{+0.60} & \textcolor{green!70!black}{+3.14}  & \textcolor{green!70!black}{1.9}  & \textcolor{green!70!black}{+3.41} & \textcolor{green!70!black}{+1.26} & \textcolor{green!70!black}{+2.67} & \textcolor{green!70!black}{146.5} &  486\\
 & SPR\_IBX   & \textcolor{green!70!black}{+3.94} & \textcolor{green!70!black}{+16.43} & \textcolor{green!70!black}{+7.02} & \textcolor{green!70!black}{0.9}  & \textcolor{green!70!black}{+1.10} & \textcolor{green!70!black}{+0.80} & \textcolor{green!70!black}{+1.00} & \textcolor{green!70!black}{172.3} &  826\\
\midrule
\multirow{6}{*}{JPG}
 & FastRoute   & 841.3k & 449.9k & 1291.2k & 6.4    & 546.6k & 385.5k & 932.0k & 1130.1 & -- \\
 & CUGR        & 815.0k & 495.2k & 1310.2k & 4.4    & 546.7k & 411.3k & 958.0k & 1179.0 & -- \\
 & SPRoute     & 848.8k & 369.8k & 1218.7k & 2.9    & 550.3k & 398.0k & 948.3k & 1048.0 & -- \\
 & FR\_JPG     & \textcolor{green!70!black}{+4.76} & \textcolor{green!70!black}{+58.78} & \textcolor{green!70!black}{+23.58} & 68.6 & \textcolor{red}{$-$0.01} & \textcolor{red}{$-$0.10} & \textcolor{red}{$-$0.05} & \textcolor{green!70!black}{309.6} & 558 \\
 & CUGR\_JPG   & \textcolor{green!70!black}{+0.76} & \textcolor{red}{$-$7.48} & \textcolor{red}{$-$2.35}                     & 4.5  & \textcolor{green!70!black}{+0.17} & \textcolor{green!70!black}{+0.36} & \textcolor{green!70!black}{+0.25} & \textcolor{green!70!black}{218.2} &  1843\\
 & SPR\_JPG    & \textcolor{green!70!black}{+6.07} & \textcolor{red}{$-$21.65} & \textcolor{red}{$-$2.34}                    & 23.7 & \textcolor{green!70!black}{+1.65} & \textcolor{green!70!black}{+1.36} & \textcolor{green!70!black}{+1.53} & \textcolor{green!70!black}{455.6} &  790\\
\midrule
\multirow{6}{*}{SWV}
 & FastRoute   & 3181.0k & 941.9k & 4122.9k & 27.3      & 2537.3k & 802.0k & 3339.3k & 1445.6 & -- \\
 & CUGR        & 3426.6k & 1017.3k & 4443.8k & 8.0       & 2799.7k & 906.8k & 3706.5k & 1289.8 & -- \\
 & SPRoute     & 3298.3k & 765.9k & 4064.2k & 7.1       & 2609.3k & 839.3k & 3448.6k & 1374.8 & -- \\
 & FR\_SWV     & \textcolor{red}{$-$2.68} & \textcolor{red}{$-$14.28} & \textcolor{red}{$-$5.33} & 40077.3 & \textcolor{red}{$-$1.34} & \textcolor{red}{$-$6.18} & \textcolor{red}{$-$2.50} & \textcolor{green!70!black}{266.5 } & 451 \\
 & CUGR\_SWV   & \textcolor{green!70!black}{+10.79} & \textcolor{green!70!black}{+19.76} & \textcolor{green!70!black}{+12.84} & \textcolor{green!70!black}{7.5} & \textcolor{green!70!black}{+8.72} & \textcolor{red}{$-$1.03} & \textcolor{green!70!black}{+6.33} & \textcolor{green!70!black}{345.1} &  739\\
 & SPR\_SWV    & \textcolor{red}{$-$3.10} & \textcolor{green!70!black}{+2.73} & \textcolor{red}{$-$2.00} & \textcolor{green!70!black}{5.8}  & \textcolor{red}{$-$0.50} & \textcolor{green!70!black}{+0.90} & \textcolor{red}{$-$0.16} & \textcolor{green!70!black}{296.4 } &  1054\\
\midrule
\multirow{6}{*}{DYN}
 & FastRoute   & 270.9k & 103.2k & 374.1k & 1.6 & 199.9k & 86.8k & 286.7k & 374.0 & -- \\
 & CUGR        & 270.8k & 111.3k & 382.1k & 1.1 & 203.6k & 94.8k & 298.4k & 291.0 & -- \\
 & SPRoute     & 274.9k & 79.9k & 354.8k & 1.1 & 202.4k & 89.1k & 291.5k & 432.1 & -- \\
 & FR\_DYN     & \textcolor{red}{$-$8.98} & \textcolor{green!70!black}{+46.61} & \textcolor{green!70!black}{+6.35} & 3.0 & \textcolor{red}{$-$0.62} & \textcolor{red}{$-$4.58} & \textcolor{red}{$-$1.82} & 503.5 & 451 \\
 & CUGR\_DYN   & \textcolor{green!70!black}{\textcolor{green!70!black}{+1.82}} & \textcolor{red}{$-$19.71} & \textcolor{red}{$-$4.45}          & 2.2 & \textcolor{green!70!black}{+2.13} & \textcolor{red}{$-$0.79} & \textcolor{green!70!black}{+1.20} & \textcolor{green!70!black}{107.5 }& 1257 \\
 & SPR\_DYN    & \textcolor{red}{$-$0.01} & \textcolor{red}{$-$0.56} & \textcolor{red}{$-$0.13}                    & \textcolor{green!70!black}{0.8} & \textcolor{green!70!black}{+0.12} & \textcolor{red}{$-$0.24} & \textcolor{green!70!black}{+0.01} & \textcolor{green!70!black}{76.1  }& 2047 \\
\midrule
\multirow{6}{*}{AR136}
 & FastRoute   & 8278.4k & 1738.1k & 10016.6k & 182.4 & 6977.7k & 1480.7k & 8458.4k & 1051.4  & -- \\
 & CUGR        & 9350.7k & 1937.5k & 11288.2k & 68.1  & 8070.1k & 1721.0k & 9791.1k & 960.0   & -- \\
 & SPRoute     & 8367.7k & 1331.5k & 9699.2k & 20.2  & 7106.4k & 1638.8k & 8745.2k & 10647.0 & -- \\
 & FR\_AR136   & \textcolor{red}{$-$0.02} & \textcolor{green!70!black}{+27.44} & \textcolor{green!70!black}{+4.74}         & 766.2 & \textcolor{green!70!black}{+0.01} & \textcolor{red}{$-$0.17} & \textcolor{red}{$-$0.02} & \textcolor{green!70!black}{971.0   }& 718 \\
 & CUGR\_AR136 & \textcolor{green!70!black}{+6.93} & \textcolor{green!70!black}{+2.99} & \textcolor{green!70!black}{+6.25} & \textcolor{green!70!black}{34.9}  & \textcolor{green!70!black}{+7.11} & \textcolor{green!70!black}{+2.47} & \textcolor{green!70!black}{+6.29} & \textcolor{green!70!black}{870.0   }& 216 \\
 & SPR\_AR136  & \textcolor{green!70!black}{+0.00} & \textcolor{green!70!black}{+0.00} & \textcolor{green!70!black}{+0.00} & \textcolor{green!70!black}{17.0}  & \textcolor{green!70!black}{+0.00} & \textcolor{green!70!black}{+0.00} & \textcolor{green!70!black}{+0.00} & \textcolor{green!70!black}{10195.0 }& 58 \\
\midrule
\multirow{6}{*}{BP}
 & FastRoute   & 7621.4k & 1781.9k & 9403.3k & 154.3 & 6325.9k & 1579.7k & 7905.5k & 951.5   & -- \\
 & CUGR        & 8111.2k & 2022.1k & 10133.3k & 43.4  & 6855.4k & 1786.3k & 8641.8k & 1400.2  & -- \\
 & SPRoute     & 7645.4k & 1369.2k & 9014.6k & 21.4  & 6385.8k & 1683.0k & 8068.8k & 34820.0 & -- \\
 & FR\_BP      & \textcolor{green!70!black}{+0.01} & \textcolor{green!70!black}{+27.42} & \textcolor{green!70!black}{+5.20} & 226.7 & \textcolor{green!70!black}{+0.01} & \textcolor{red}{$-$0.02} & \textcolor{green!70!black}{+0.00} & \textcolor{green!70!black}{641.3   }& 643 \\
 & CUGR\_BP    & \textcolor{green!70!black}{+2.28} & \textcolor{green!70!black}{+0.81} & \textcolor{green!70!black}{+1.98}  & \textcolor{green!70!black}{38.8}  & \textcolor{green!70!black}{+2.15} & \textcolor{green!70!black}{+0.37} & \textcolor{green!70!black}{+1.78} & \textcolor{green!70!black}{927.3   }& 185 \\
 & SPR\_BP     & \textcolor{green!70!black}{+0.00} & \textcolor{green!70!black}{+0.00} & \textcolor{green!70!black}{+0.00}  & \textcolor{green!70!black}{13.7}  & \textcolor{green!70!black}{+0.01} & \textcolor{green!70!black}{+0.02} & \textcolor{green!70!black}{+0.00} & \textcolor{green!70!black}{25095.0 }& 58 \\
\bottomrule
\end{tabular}
}%
\vspace{-8mm}
\end{table}

\begin{table}[t]
\centering
\caption{QoR Metrics post GR and post DR of selected evolved router for SKY130HD. Positive $\Delta$WL denotes reduction.}
\vspace{-2mm}
\label{tab:sky130_dr_wl_gr_rt}
\scriptsize
\setlength{\tabcolsep}{2.5pt}
\resizebox{\columnwidth}{!}{%
\begin{tabular}{l|l|cccc|cccc|c}
\toprule
 & & \multicolumn{4}{c|}{\textbf{GR}} & \multicolumn{4}{c|}{\textbf{DR}} & \\
 & & WL($\mu$m) & VC & TWL($\mu$m) & & WL($\mu$m) & VC & TWL($\mu$m) & & \\
\textbf{Des.} & \textbf{Router} & $\Delta$WL & $\Delta$VC & $\Delta$TWL & RT(s) & $\Delta$WL & $\Delta$VC & $\Delta$TWL & RT(s) & \textbf{\shortstack{\#LoCM}}\\
\midrule
\multirow{6}{*}{AES}
 & FastRoute   & 1079.2k & 127.7k & 1206.9k & 5.3  & 780.2k & 123.0k & 903.2k & 318.5 & -- \\
 & CUGR        & 1199.6k & 126.1k & 1325.7k & 2.1  & 836.8k & 126.6k & 963.4k & 321.9 & -- \\
 & SPRoute     & 1096.2k & 103.1k & 1199.3k & 1.4  & 780.7k & 126.8k & 907.5k & 351.2 & -- \\
 & FR\_AES     & \textcolor{green!70!black}{+2.00}  & \textcolor{green!70!black}{+5.07}  & \textcolor{green!70!black}{+2.32}  & 8.8 & \textcolor{green!70!black}{+0.82}  & \textcolor{red}{$-$3.11}                  & \textcolor{green!70!black}{+0.28}  & 500.7  & 2809 \\
 & CUGR\_AES   & \textcolor{green!70!black}{+15.29} & \textcolor{red}{$-$20.54}          & \textcolor{green!70!black}{+15.79} & \textcolor{green!70!black}{1.9}  & \textcolor{green!70!black}{+8.56}  & \textcolor{red}{$-$0.88} & \textcolor{green!70!black}{+7.32}  & \textcolor{green!70!black}{319.4}  & 276\\
 & SPR\_AES    & \textcolor{green!70!black}{+4.68}  & \textcolor{red}{$-$13.56}          & \textcolor{green!70!black}{+2.64}  & \textcolor{green!70!black}{0.8}  & \textcolor{green!70!black}{+1.82}  & \textcolor{red}{$-$1.90} & \textcolor{green!70!black}{+1.30}  & 392.7  & 528\\
\midrule
\multirow{6}{*}{IBX}
 & FastRoute   & 921.8k & 123.0k & 1044.8k & 5.0  & 657.9k & 132.9k & 790.8k & 1263.0 & -- \\
 & CUGR        & 966.6k & 120.1k & 1086.7k & 1.4  & 675.0k & 134.7k & 809.7k & 1258.3 & -- \\
 & SPRoute     & 1028.4k & 106.7k & 1135.1k & 1.8  & 700.1k & 138.5k & 838.6k & 1593.4 & -- \\
 & FR\_IBX    & \textcolor{red}{$-$18.42}           & \textcolor{green!70!black}{+29.55} & \textcolor{red}{$-$12.77}  & \textcolor{green!70!black}{3.5}  & \textcolor{green!70!black}{+1.83}          & \textcolor{green!70!black}{+2.82}          & \textcolor{green!70!black}{+2.00}  & \textcolor{green!70!black}{669.5} & 1850\\
 & CUGR\_IBX  & \textcolor{green!70!black}{+7.41}   & \textcolor{green!70!black}{+20.71} & \textcolor{green!70!black}{+8.87}  & \textcolor{green!70!black}{1.1}  & \textcolor{green!70!black}{+3.91} & \textcolor{green!70!black}{+3.47} & \textcolor{green!70!black}{+3.84}  & \textcolor{green!70!black}{444.5} & 456\\
 & SPR\_IBX   & \textcolor{green!70!black}{+2.18}  & \textcolor{green!70!black}{+3.37}  & \textcolor{green!70!black}{+2.29}  & \textcolor{green!70!black}{1.4}  & \textcolor{green!70!black}{+2.39} & \textcolor{green!70!black}{+3.96} & \textcolor{green!70!black}{+2.65}  & \textcolor{green!70!black}{622.5} & 60 \\
\midrule
\multirow{6}{*}{JPG}
 & FastRoute   & 1888.7k & 296.5k & 2185.1k & 11.5 & 1254.2k & 287.3k & 1541.5k & 943.5 & -- \\
 & CUGR        & 1853.3k & 285.1k & 2138.3k & 4.8  & 1249.8k & 288.3k & 1538.2k & 891.2 & -- \\
 & SPRoute     & 1882.7k & 225.1k & 2107.8k & 4.3  & 1246.3k & 292.1k & 1538.3k & 868.2 & -- \\
 & FR\_JPG    & \textcolor{red}{$-$0.98}          & \textcolor{green!70!black}{+28.60} & \textcolor{green!70!black}{+3.04}  & \textcolor{green!70!black}{3.5}  & \textcolor{green!70!black}{+2.51} & \textcolor{green!70!black}{+0.62}          & \textcolor{green!70!black}{+2.16}  & 1194.5 & 1268\\
 & CUGR\_JPG  & \textcolor{green!70!black}{+0.84} & \textcolor{red}{$-$0.79}  & \textcolor{green!70!black}{+3.62}  & 4.8                              & \textcolor{green!70!black}{+0.80} & \textcolor{green!70!black}{+0.63} & \textcolor{green!70!black}{+0.77}  & 5315.0 & 225\\
 & SPR\_JPG   & \textcolor{green!70!black}{+2.38} & \textcolor{red}{$-$43.13} & \textcolor{green!70!black}{+3.48}  & \textcolor{green!70!black}{2.5}  & \textcolor{green!70!black}{+0.38} & \textcolor{red}{$-$0.17}          & \textcolor{green!70!black}{+0.28}  & 1118.0 & 64\\
\bottomrule
\end{tabular}
}
\vspace{-8mm}
\end{table}

\subsubsection{Wirelength and via counts}

Across all three technology nodes, we observe improvements in DR WL for a majority of design–router pairs. For example, in ASAP7 (Table~\ref{tab:asap7_dr_wl_gr_rt}), FR\_AES achieves a $+1.09$\% improvement in DR WL with a negligible change in DR VC ($-0.07$\%), while SPR\_IBEX achieves a $+1.02$\% DR WL improvement with a $+0.83$\% improvement in DR VC. Similarly, in Nangate45 (Table~\ref{tab:nangate45_dr_wl_gr_rt}),  CUGR\_SWERV achieves a significant improvement of $+8.72$\% in DR WL with only a $-1.03$\% change in DR VC, demonstrating strong gains in routing quality. In SKY130HD (Table~\ref{tab:sky130_dr_wl_gr_rt}), CUGR\_AES achieves a $+8.56$\% improvement in DR WL with a $-0.88$\% degradation in DR VC, indicating that substantial WL improvement can be achieved alongside stable or improved via counts. At the GR stage, improvements are more variable and do not always translate to DR gains. For instance, in Nangate45, FR\_JPEG shows a $+4.76$\% improvement in GR WL and a large $+58.78$\% improvement in GR VC, yet results in a slight $-0.01$\% degradation in DR WL and $-0.10$\% in DR VC. In contrast, SPR\_IBEX (Nangate45) shows  $+3.94$\% GR WL improvement as well as $+1.10$\% DR WL improvement, highlighting that improvements in GR metrics do not always translate consistently to DR outcomes. Overall, these results reinforce that DR WL is the most reliable indicator of final routing quality.

\subsubsection{Scale of the Diff}
The final column in each table reports the number of lines of code modified (\#LoCM) during evolution relative to the baseline router. We observe that significant QoR improvements can be achieved with relatively small code changes in some cases, while other designs require larger modifications. The relationship between design size and modification scale is not monotonic; both small and large designs may require substantial or minimal changes depending on structure. For example, Table~\ref{tab:nangate45_dr_wl_gr_rt} shows that CUGR\_AES achieves only a +2.21\% improvement in DR WL with 5341 lines of code changes, whereas CUGR\_AR136 requires only 216 lines of modifications to achieve +7.11\% in DR WL improvement. This indicates that GR-Evolve can discover both fine-grained heuristic adjustments and larger structural modifications to routing algorithms, depending on the design characteristics.

\begin{table*}[t]
\centering
\caption{Comparison of \textbf{CUGR\_SWERV} on Nangate45 against baseline global routers on ICCAD19 benchmarks~\cite{iccad19-contest}. Positive $\Delta$WL  denote reduction relative to the baseline, computed as $(\text{baseline} - \text{ours})/\text{baseline} \times 100$.}
\vspace{-4mm}
\label{tab:iccad19_combined}
\scriptsize
\setlength{\tabcolsep}{2.5pt}
\begin{tabular}{l|ccccc|ccccc|ccccc|ccccc}
\toprule
\multirow{2}{*}{\textbf{Benchmark}} &
\multicolumn{5}{c|}{\textbf{Ours (CUGR\_SWERV)}} &
\multicolumn{5}{c|}{\textbf{CUGR}} &
\multicolumn{5}{c|}{\textbf{FastRoute}} &
\multicolumn{5}{c}{\textbf{SPRoute}} \\
& MO & TO & WL & VC & RT(s)
& MO & TO & $\Delta$WL & $\Delta$VC & RT(s)
& MO & TO & $\Delta$WL & $\Delta$VC & RT(s)
& MO & TO & $\Delta$WL & $\Delta$VC & RT(s) \\
\midrule
\texttt{ispd19\_test1}
& 0 & 0 & 85036 & 37499 & 4.1
& 0 & 0 & \textcolor{green!70!black}{+0.41\%} & \textcolor{green!70!black}{+2.65\%} & 0.6
& 0 & 0 & \textcolor{green!70!black}{+1.99\%} & \textcolor{green!70!black}{+6.23\%} & 1.0
& 0 & 0 & \textcolor{green!70!black}{+1.94\%} & -- & 0.4 \\
\texttt{ispd19\_test2}
& 0 & 0 & 2945848 & 809935 & 27.1
& 1 & 267 & \textcolor{green!70!black}{+0.06\%} & \textcolor{red}{$-$0.06\%} & 6.2
& 0 & 0 & \textcolor{green!70!black}{+2.07\%} & \textcolor{green!70!black}{+6.46\%} & 28.4
& 2 & 8 & \textcolor{green!70!black}{+1.69\%} & -- & 4.8 \\
\texttt{ispd19\_test3}
& 0 & 0 & 115479 & 64455 & 3.2
& 2 & 250 & \textcolor{green!70!black}{+0.12\%} & \textcolor{green!70!black}{+1.17\%} & 0.7
& 0 & 0 & \textcolor{green!70!black}{+4.75\%} & \textcolor{green!70!black}{+8.61\%} & 0.8
& 0 & 0 & \textcolor{green!70!black}{+3.56\%} & -- & 0.5 \\
\texttt{ispd19\_test4}
& 0 & 0 & 5845119 & 760311 & 20.4
& 1 & 2274 & \textcolor{green!70!black}{+16.43\%} & \textcolor{green!70!black}{+3.50\%} & 11.6
& 0 & 0 & \textcolor{green!70!black}{+6.92\%} & \textcolor{red}{$-$127.41\%} & 27.0
& -- & -- & -- & -- & -- \\
\texttt{ispd19\_test6}
& 0 & 0 & 7771972 & 2020881 & 67.6
& 1 & 1172 & \textcolor{green!70!black}{+0.23\%} & \textcolor{green!70!black}{+1.44\%} & 17.3
& 0 & 0 & \textcolor{green!70!black}{+2.07\%} & \textcolor{green!70!black}{+6.71\%} & 79.2
& 15 & 13117 & \textcolor{green!70!black}{+1.75\%} & -- & 37.0 \\
\texttt{ispd19\_test7}
& 0 & 0 & 14402101 & 3848791 & 108.7
& 1 & 1644 & \textcolor{green!70!black}{+0.35\%} & \textcolor{red}{$-$0.52\%} & 36.2
& 0 & 0 & \textcolor{green!70!black}{+1.84\%} & \textcolor{green!70!black}{+8.74\%} & 143.2
& 15 & 9289 & \textcolor{green!70!black}{+1.58\%} & -- & 43.8 \\
\texttt{ispd19\_test8}
& 0 & 0 & 22187136 & 6117798 & 146.1
& 0 & 1146 & \textcolor{green!70!black}{+0.06\%} & \textcolor{green!70!black}{+2.99\%} & 49.1
& 0 & 0 & \textcolor{green!70!black}{+2.01\%} & \textcolor{green!70!black}{+5.46\%} & 227.4
& 15 & 2221 & \textcolor{green!70!black}{+1.61\%} & -- & 41.0 \\
\texttt{ispd19\_test9}
& 0 & 0 & 34047838 & 10142672 & 198.6
& 2 & 1157 & \textcolor{green!70!black}{+0.01\%} & \textcolor{green!70!black}{+3.40\%} & 80.1
& 0 & 0 & \textcolor{green!70!black}{+2.18\%} & \textcolor{green!70!black}{+5.29\%} & 367.3
& 15 & 3428 & \textcolor{green!70!black}{+1.78\%} & -- & 71.9 \\
\texttt{ispd19\_test10}
& 0 & 0 & 33592551 & 9663225 & 200.4
& 2 & 5316 & \textcolor{red}{$-$0.02\%} & \textcolor{red}{$-$0.02\%} & 76.0
& 0 & 0 & \textcolor{green!70!black}{+2.58\%} & \textcolor{green!70!black}{+8.99\%} & 383.4
& 15 & 6462 & \textcolor{green!70!black}{+1.73\%} & -- & 65.9 \\
\bottomrule
\end{tabular}
\vspace{-4mm}
\end{table*}

\subsubsection{Runtimes}

We report the runtime of the evolved global routers and the corresponding DR runtime in Tables~\ref{tab:asap7_dr_wl_gr_rt}, ~\ref{tab:nangate45_dr_wl_gr_rt}, and~\ref{tab:sky130_dr_wl_gr_rt} for all benchmarks considered. Since the GR runtime is the lowest-priority objective in the optimization hierarchy and routers are selected based on the best post-DR WL from the explored set, the resulting runtimes are often higher than those of baseline routers, though some evolved routers achieve comparable or improved runtime.

Alternative selections prioritizing runtime could be made if desired, as the explored set contains routers with different trade-offs shown earlier. There are routers that also provide better WL with improved runtime. 
We report absolute runtime values for clarity, as percentage degradation or improvements on small runtimes (few seconds) are not very meaningful. In some cases, improved global routing solutions lead to faster convergence during DR iterations.

We also report the runtime of the evolution process. For designs with fewer than 20{,}000 nets, we run 75 evolution iterations, which takes approximately 24 hours. This is a one-time cost incurred per design. For designs with more than 20{,}000 nets, we use warm-started evolution and run 25 iterations (e.g., AR136 and BP), resulting in a runtime of approximately 12--18 hours. Since this cost is incurred only once per design, it is amortized over subsequent uses of the evolved router and is therefore less critical compared to per-run routing runtime reported in Tables~\ref{tab:asap7_dr_wl_gr_rt},~\ref{tab:nangate45_dr_wl_gr_rt}, and~\ref{tab:sky130_dr_wl_gr_rt}.

\subsubsection{ICCAD19 Contest Benchmarks Comparisons QoR}
\noindent
While, in principle, evolution could be performed for each router on these designs, the other contest routers are static. We therefore evaluate a single evolved router, CUGR\_SWERV (evolved on NanGate45), on the ICCAD19 contest benchmarks. These benchmarks are executed in a standalone setting and are not integrated with ORFS, as they already provide placed DEFs and use a different technology node. The post-GR results are summarized in Table~\ref{tab:iccad19_combined}. We report full raw metrics for CUGR\_SWERV. For the baselines, we report maximum overflow (MO) and total overflow (TO), along with percentage changes of CUGR\_SWERV relative to each baseline for WL, VC, and RT. Note that baselines are not re-optimized per design, unlike the evolved router, which may bias results. Positive $\Delta$WL and $\Delta$VC indicate improvement, and we also report absolute runtime values. CUGR\_SWERV generally reduces WL and, in many cases, VC relative to CUGR and FastRoute, while achieving zero overflow, satisfying routing feasibility constraints across all evaluated benchmarks. SPRoute  does not report VC in its standalone format for these benchmarks; in prior experiments, VC was obtained through OpenROAD-based evaluation. Since the selected router is optimized primarily for DR WL, with runtime treated as a lower-priority objective, its runtime is generally higher than CUGR and SPRoute. However, it has better runtimes than FastRoute. SPRoute did not complete execution on \texttt{ispd19\_test4}.

\begin{figure}[t]
\centering
\vspace{-4pt}
\begin{subfigure}[t]{0.44\columnwidth}
\begin{lstlisting}[style=diffleft, title={\scriptsize\textbf{CUGR (Base)}}, firstnumber=120]
GridGraphView<CostT> wireCostView;
grid_graph_->extractWireCostView(
  wireCostView);
sortNetIndices(netIndices);
|\colorbox{diffred}{\strut\color{diffredtxt}SparseGrid grid(10, 10, 0, 0);}|
for (const int netIndex
       : netIndices) {
  GRNet* net =
    gr_nets_[netIndex].get();
  MazeRoute mazeRoute(net,
    grid_graph_.get(), logger_);
  mazeRoute
    .constructSparsifiedGraph(
      wireCostView, grid);
\end{lstlisting}
\end{subfigure}%
\hfill
\begin{subfigure}[t]{0.48\columnwidth}
\begin{lstlisting}[style=diffright, title={\scriptsize\textbf{CUGR\_AES (Evolved)}}]
for (const int netIndex
       : netIndices) {
  GRNet* net =
    gr_nets_[netIndex].get();
|\colorbox{diffgreen}{\strut\color{diffgreentxt}  const int hp =}|
|\colorbox{diffgreen}{\strut\color{diffgreentxt}    std::max(net->getBoundingBox()}|
|\colorbox{diffgreen}{\strut\color{diffgreentxt}      .hp(), 1);}|
|\colorbox{diffgreen}{\strut\color{diffgreentxt}  std::vector<CandidateGrid>}|
|\colorbox{diffgreen}{\strut\color{diffgreentxt}    candidateGrids;}|
|\colorbox{diffgreen}{\strut\color{diffgreentxt}  pushGridCandidate(...);}|
|\colorbox{diffgreen}{\strut\color{diffgreentxt}  pushGridCandidate(...);}|
|\colorbox{diffgreen}{\strut\color{diffgreentxt}  for (const auto\& candidate}|
|\colorbox{diffgreen}{\strut\color{diffgreentxt}         : candidateGrids) \{}|
|\colorbox{diffgreen}{\strut\color{diffgreentxt}    SparseGrid grid(...);}|
|\colorbox{diffgreen}{\strut\color{diffgreentxt}    // route and score tree}|
|\colorbox{diffgreen}{\strut\color{diffgreentxt}    if (candidateScore.objective}|
|\colorbox{diffgreen}{\strut\color{diffgreentxt}        < bestScore.objective)}|
|\colorbox{diffgreen}{\strut\color{diffgreentxt}      bestTree = candidateTree;}|
|\colorbox{diffgreen}{\strut\color{diffgreentxt}  \}}|
}
\end{lstlisting}
\end{subfigure}%
\vspace{-4pt}
\caption{Base CUGR uses one sparse-grid configuration, while \texttt{CUGR\_AES} evaluates multiple candidate grids per net and keeps the best-scoring route tree.}
\label{fig:newgr-grid}
\vspace{-16pt}
\end{figure}

\section{Code Diffs Discussion}\label{sec: discussion_section}

We highlight one representative evolved mutation of \texttt{CUGR\_AES} on Nangate45 and analyze the algorithmic changes in relation to routing behavior. We examine how the evolved implementation modifies the baseline CUGR source code in terms of pipeline, topology construction, search strategy, and congestion handling. This case study shows that the evolved router captures meaningful routing principles rather than simply retuning parameters. Furthermore, analysis of code diffs and reasoning traces provides insights into both algorithmic changes and underlying design characteristics.

\texttt{CUGR\_AES} modifies both route generation and evaluation, and introduces post-processing stages that restructure routing solutions. Instead of a single sparse grid per net, it generates four candidate grids at different resolutions, with spacing adapted to net half-perimeter. Each grid is offset, fully routed, and scored based on WL, VC, and overflow, with the best candidate selected. This code difference is shown in Fig.~\ref{fig:newgr-grid} as an example generated mutation. For multi-pin nets, \texttt{CUGR\_AES} replaces FLUTE with a two-hub star topology. The two farthest pins act as hubs, remaining pins attach to the nearest hub, and shared bends promote trunk reuse. The most significant changes are three post-processing stages after pattern routing. First, a wirelength pulse stage iteratively reroutes high-impact nets using multiple candidate grids. Second, a global rebalance stage rips up all nets and reroutes from scratch, removing path dependencies of sequential routing and exploring different orderings. Third, a compaction stage targets long multi-pin nets with finer grids to reduce WL.  Together, these changes yield more compact global routes, resulting in lower WL after DR.

\section{Conclusion}\label{sec: conclusion}
We presented GR-Evolve, an agentic framework for design-adaptive global routing through LLM-driven code evolution. The framework operates by exploring a space of router variants enabling design--tool co-exploration. GR-Evolve structures this process through a stateless execution model, persistent context, and a controlled mutation pipeline that allows the agent to systematically generate and evaluate diverse routing strategies. Rather than relying on fixed heuristics or manual tuning, the framework enables automated generation of candidate routers tailored to individual designs. This approach represents a paradigm whose practical feasibility has been significantly expanded by the emergence of LLMs: the ability to autonomously reason about, modify, and evolve complex EDA tool implementations at scale.

\clearpage
\bibliographystyle{ieeetr}
\bibliography{references}

\end{document}